\documentstyle[12pt]{article}
\newcommand{\cl}{\centerline}
\renewcommand{\theequation}{\arabic{equation}}
\newcommand{\beq}{\begin{equation}}
\newcommand{\eeq}{\end{equation}}
\newcommand{\bea}{\begin{eqnarray}}
\newcommand{\eea}{\end{eqnarray}}
\newcommand{\X}{{\vec X}}
\newcommand{\pro}{\partial}
\newcommand{\n}{\hat n}
\newcommand{\hn}{\hat{n}}
\newcommand{\oneg}{\displaystyle\frac{1}{g}}

\newcommand{\D}{{\hat D}}

\newcommand{\A}{{\vec A}}
\newcommand{\valpha}{{\vec \alpha}}

\newcommand{\dfrac}{\displaystyle\frac}
\newcommand{\ba}{\begin{array}}
\newcommand{\ea}{\end{array}}

\newcommand{\nn}{\nonumber}

\begin{document}

\begin{titlepage}
\setlength{\textwidth}{5.0in} \setlength{\textheight}{7.5in}
\setlength{\parskip}{0.0in} \setlength{\baselineskip}{18.2pt}
\begin{center}
{\Large{{\bf Dirac Quantization of Restricted QCD}}}\par
\vskip0.5cm
\begin{center}
{Y.M. Cho$^{1,a}$, Soon-Tae Hong$^{2,b}$, J.H. Kim$^{1}$, and
Young-Jai Park$^{3,c}$}\par
\end{center}
\vskip 0.4cm
\begin{center}
{$^{1}$Center for Theoretical Physics and School of Physics}\par
{College of Natural Sciences, Seoul National University, Seoul
151-742, Korea}\par {$^{2}$Department of Science Education and
Research Institute for Basic Sciences}\par{Ewha Womans University,
Seoul 120-750 Korea}\par {$^{3}$Department of Physics and Basic
Science Research Institute,}\par {Sogang University, C.P.O. Box
1142, Seoul 100-611, Korea}\par
\end{center}
\vskip 0.3cm
\cl{\today}
\vfill
\begin{center}
{\bf ABSTRACT}
\end{center}
\begin{quotation}
We discuss the quantization of the restricted gauge theory of
SU(2) QCD regarding it as a second-class constraint system,
and construct the BRST symmetry of the constrained
system in the framework of the improved Dirac quantization scheme.
Our analysis tells that one could quantize the restricted QCD
as if it is a first-class constraint system.
\vskip 0.5cm
\noindent
---------------------------------------------------------------------\\
{\small $^{a}$ymcho@yongmin.snu.ac.kr;
$^{b}$soonhong@ewha.ac.kr;$^{c}$yjpark@ccs.sogang.ac.kr}\par
\noindent \vskip 0.5cm \noindent
\end{quotation}
\end{center}
\end{titlepage}

\newpage


\section{Introduction}
\setcounter{equation}{0}
\renewcommand{\theequation}{\arabic{section}.\arabic{equation}}

Understanding QCD and its confinement mechanism is probably
one of the most fundamental problems in theoretical physics.
It has long been speculated that the confinement could be explained
by the condensation of the non-Abelian monopoles which
could trigger the dual Meissner effect \cite{nambu,cho1,hooft,cho2}.
In particular, it has been argued that the dual dynamics
of the restricted QCD, restricted by the ``Abelian projection''
of the QCD potential, plays the important role
in the confinement of color \cite{cho1,hooft,cho00}.
But understanding the dual dynamics of the restricted QCD
which describes the interaction between the chromoelectric gluon
and the chromomagnetic monopole has been non-trivial,
because it describes
the dynamics of a constrained system. The purpose of this paper
is to clarify the dual dynamics and quantize
the restricted QCD using the improved
Dirac quantization method \cite{dirac,BFT,pr01}.

The proof of the monopole condensation in QCD has
been extremely difficult to achieve.
The earlier attempts to prove the
magnetic condensation with the effective action of QCD has failed
because the external magnetic background is unstable
due to the tachyonic modes \cite{sav,yil}. Recently, however,
there has been interesting new developments on this old problem.
First, it has been shown that the instability of the magnetic background
follows from the fact that the external background is not gauge invariant.
Indeed, if we separate the magnetic background gauge independently
from the quantum fluctuation of gluon and impose the gauge invariance to
the magnetic background or the causality to
the gluon fluctuation around the magnetic background,
we can show from the resulting effective action of QCD
that the magnetic background becomes a stable
vacuum of QCD \cite{prd02,jhep}.

Independent of this it has been conjectured that
the Skyrme-Faddeev Lagrangian which allows knot-like topological
solitons could be interpreted as an effective Lagrangian of QCD
in the low energy limit \cite{faddeev1,faddeev2}.
Although this conjecture may not be completely correct,
it does reveal an interesting connection
between the Skyrme-Faddeev theory and QCD, at least in
some approximation \cite{niemi99,cho01}.
The two theories have almost identical topological structures.
Both admit the Wu-Yang monopole characterized by
the mapping $\pi_2(S^2)$ as a fundamental object,
and allow the knot configurations classified
by the mapping $\pi_3(S^3) \simeq \pi_3(S^2)$ as classical
solutions \cite{prl01,plb05}. Besides, the knot of Skyrme-Faddeev
theory can be mapped to QCD and interpreted to describe the
multiple vacua of QCD \cite{plb06}. Moreover, both can be
interpreted as theories of confinement.
The Skyrme-Faddeev theory screens (confines) the magnetic flux of the
monopole-antimonopole pairs, while QCD confines
the chromoelectric flux of the gluon-antigluon pairs.
Furthermore, one can show that one may actually
derive a generalized Skyrme-Faddeev action from
the QCD effective action itself \cite{cho01}.

The crux behind these new developments is the so-called
``Cho decomposition'', the gauge
independent decomposition of the non-Abelian potential into the
restricted potential and the valence potential \cite{cho1,cho2}.  The
virtue of the decomposition is that it clarifies the topological
structure of the non-Abelian gauge theory, and naturally takes
care of the topological characters in the dynamics.  It has been
well-known that the non-Abelian gauge theory has rich topological
structures manifested by the non-Abelian monopoles
\cite{wu,choprl80}, the multiple vacua and the
instantons~\cite{bpst,cho79}.  One has to make sure that these
topological characters are properly taken into account in the
non-Abelian dynamics. In the specific case of SU(2), the relevant
topology is the $\pi_2 (S^2)$ which describes the non-Abelian
monopoles, and the $\pi_3(S^3)$ which describes the multiple vacua
and the corresponding vacuum tunneling by instantons. Since the
decomposition of the non-Abelian connection contains these
topological degrees explicitly, it can naturally take care of them
in the non-Abelian dynamics \cite{cho1,cho2}.

An important consequence of the decomposition is that it allows us
to view QCD as the restricted gauge theory (made of the restricted
potential) which is coupled to a gauge-covariant colored vector
field (the valence potential).  The restricted potential is
defined in such a way that it allows a covariantly constant unit
isovector $\hat n$ everywhere in space-time, which enables us to
define the gauge-independent color direction everywhere in
space-time and at the same time allows us to define the magnetic
potential of the non-Abelian monopoles. Furthermore it has the full
SU(2) gauge degrees of freedom, in spite of the fact that it is
restricted \cite{cho1,cho2}. Consequently
the restricted QCD made of the restricted
potential describes a very interesting dual dynamics of its own,
and plays a crucial role in our understanding of QCD
\cite{cho00,cho01}.

On the other hand the restricted QCD is a constrained system, due
to the presence of the topological field $\hat n$ which is
constrained to have the unit norm. Physical system with
constraints was systematically studied by Dirac~\cite{dirac}, who
showed that in the second-class constraint system one needs to use
the Dirac brackets instead of the Poisson brackets to quantize the
system.  The standard Dirac method was later improved to convert
the second-class constraints into first-class ones with the
introduction of St\"uckelberg fields~\cite{BFT,pr01}. Recently
this improved Dirac quantization scheme has been applied to
several areas of current interests such as the soliton models
\cite{sk2,o3,hong05prd}, high dense matter physics~\cite{hong00q},
and D-brane systems \cite{hong00db}.

In this paper we discuss the second-class
constraints of the restricted QCD and quantize the theory in
the framework of the improved Dirac quantization method. In
specific, we first quantize the constrained system and obtain the
first-class effective Lagrangian, and discuss the
Becci-Rouet-Stora-Tyutin (BRST) symmetry arising
from the first-class effective Lagrangian. Our analysis tells that
the restricted QCD, even though it is a second-class constraint
system, could actually be quantized as a first-class constraint
system.

The paper is organized as follows. In Section 2, we briefly
recapitulate the restricted QCD. In Section 3, we analyze the
constraint structure of the theory.  In Section 4, we extend this
theory to the first-class Hamiltonian system using the improved
Dirac quantization scheme and obtain the first-class Lagrangian
using the Faddeev-Senjanovic formula. In Section 5, we construct
the BRST symmetry of the restricted QCD after obtaining the
first-class effective Lagrangian. Finally in Section 6 we discuss
the physical implications of our work.


\section{Restricted QCD}
\setcounter{equation}{0}
\renewcommand{\theequation}{\arabic{section}.\arabic{equation}}

In this section we briefly review the restricted QCD.  Consider
SU(2) gauge theory for simplicity.  A natural way to accommodate
the topological degrees into the theory is to introduce a topological
field $\n$ of unit length, and to decompose the
connection into the Abelian projection part which leaves $\n$ a
covariant constant and the remaining part which forms a covariant
vector field~\cite{cho1,cho2},
\bea \vec{A}_\mu = A_\mu \n -
\oneg \n\times\pro_\mu\n+\X_\mu
            = \hat A_\mu + \X_\mu, \nn\\
A_\mu = \n\cdot \vec A_\mu,~~~\n^2 =1, \label{amu}~~~~~~~~~~~~
\eea
where $A_\mu$ is the ``electric'' potential. Notice that the
Abelian projection $\hat A_\mu$ is precisely the connection which
leaves $\n$ invariant under the parallel transport and makes $\n$
a covariant constant,
\bea \D_\mu \n
= \pro_\mu \n + g {\hat A}_\mu \times \n = 0.
\eea
Under the infinitesimal gauge transformation
\bea \delta \n = - \vec
\alpha\times\n,~~~\delta \A_\mu = \oneg  D_\mu \vec \alpha,
\eea
one has
\bea \delta A_\mu = \oneg \n \cdot \pro_\mu\valpha,~~~
\delta \hat A_\mu = \oneg \D_\mu \valpha,~~~ \delta \X_\mu =
-\valpha \times \X_\mu.
\eea
This shows that $\hat A_\mu$ by itself describes an SU(2)
connection which enjoys the full SU(2) gauge degrees of freedom.

We emphasize that the crucial element in our decomposition
(\ref{amu}) is the restricted potential $\hat A_\mu$
defined by the Abelian projection. Once this
part is identified, the expression follows immediately due to the
fact that the connection space (the space of all gauge potentials)
forms an affine space \cite{cho1,cho2}. Indeed the affine nature
of the connection space guarantees that one can describe an
arbitrary potential simply by adding a gauge-covariant piece $\vec
X_\mu$ to the restricted potential. The decomposition (\ref{amu}),
which has recently become known as the ``Cho decomposition''
\cite{faddeev2} or the ``Cho-Faddeev-Niemi decomposition''
\cite{niemi99}, was introduced long time ago in an attempt to
demonstrate the monopole condensation in QCD~\cite{cho1,cho2}.
However, only recently the importance of the decomposition in
clarifying the non-Abelian dynamics has become appreciated by many
authors~\cite{faddeev2,niemi99,zucc,kondo}. Indeed this decomposition
has played a crucial role for us to establish the ``Abelian dominance''
in Wilson loops in QCD \cite{cho00}, and to calculate the effective
action of QCD to prove the monopole condensation \cite{prd02,jhep,kondo}.

To understand the physical meaning of our decomposition notice
that the restricted potential $\hat{A}_\mu$ actually has a dual
structure. Indeed the field strength made of the restricted
potential is decomposed as
\begin{eqnarray}
\hat{F}_{\mu\nu}=(F_{\mu\nu}+ H_{\mu\nu})\hat{n}\mbox{,}~~~~~~~~~~~~~~ \nonumber\\
F_{\mu\nu}=\partial_\mu A_{\nu}-\partial_{\nu}A_\mu,~~~~~~~~~~~~~~ \nonumber\\
H_{\mu\nu}=-\dfrac{1}{g}
\hat{n}\cdot(\partial_\mu\hat{n}\times\partial_\nu\hat{n})
=\partial_\mu \tilde C_\nu-\partial_\nu \tilde C_\mu,
\end{eqnarray}
where $\tilde C_\mu$ is the ``magnetic'' potential
\cite{cho1,cho2}. Notice that one can always introduce the
magnetic potential (at least locally section-wise), since one has
the following identity
\bea
\partial_\mu {\tilde H}_{\mu\nu} = 0,~~~{\tilde H}_{\mu\nu} =
\dfrac{1}{2} \epsilon_{\mu\nu\rho\sigma} H_{\rho\sigma}.
\eea
This allows us to  identify the non-Abelian monopole potential by
\bea \vec C_\mu= -\frac{1}{g}\hat n \times \partial_\mu\hat n,
\eea
in terms of which the magnetic field is expressed as
\bea
\vec H_{\mu\nu}=\partial_\mu \vec C_\nu-\partial_\nu \vec C_\mu+ g
\vec C_\mu \times \vec C_\nu =-\frac{1}{g}
\partial_\mu\hat{n}\times\partial_\nu\hat{n}=H_{\mu\nu}\hat n.
\eea

Another important feature of $\hat{A}_\mu$ is that, as an SU(2)
potential, it retains the full topological characteristics of the
original non-Abelian potential \cite{cho1,cho2}.
Clearly the isolated singularities
of $\hat{n}$ defines $\pi_2(S^2)$ which describes the non-Abelian
monopoles.  Indeed $\hat A_\mu$ with $A_\mu =0$ and $\hat n= \hat
r$ (or equivalently, $\vec C_\mu$ with $\hat n= \hat r$) describes
precisely the Wu-Yang monopole \cite{wu,choprl80}.  Besides, with
the $S^3$ compactification of $R^3$, $\hat{n}$ characterizes the
Hopf invariant $\pi_3(S^2)\simeq\pi_3(S^3)$ which describes the
topologically distinct vacua \cite{bpst,cho79}. This tells that
the restricted gauge theory made of $\hat A_\mu$ could describe
the dual dynamics which should play an essential role in SU(2)
QCD, which displays the full topological characters of the
non-Abelian gauge theory~\cite{cho00,cho01}.

With the connection (\ref{amu}), we have
\bea
\vec{F}_{\mu\nu}=(F_{\mu\nu}+ H_{\mu\nu})\hat{n}+\D _\mu
\X_\nu-\D_\nu \X_\mu
+g\X_\mu \times \X_\nu, \nn \\
\hn \cdot \X_\mu = 0, ~~~\hn \cdot \hat D_\mu \X_\nu =
0,~~~~~~~~~~~~~~~~~~
\eea
so that the Yang-Mills Lagrangian density is expressed as
\bea
{\cal L}=-\frac{1}{4}\hat{F}_{\mu\nu}^2 -\frac{1}{4} (\D_\mu
\X_\nu -\D_\nu \X_\mu)^2
-\frac{g}{2} {\hat F}_{\mu\nu}\cdot (\X_\mu \times \X_\nu) \nonumber\\
-\frac{g^{2}}{4}(\X_\mu \times \X_\nu)^2 +\lambda(\hat n^2
-1)+\lambda_\mu \hat n \cdot \X_\mu,~~~~~~~~~
\eea
where $\lambda$ and $\lambda_\mu$ are the Lagrangian multipliers.

Now we consider the restricted gauge theory made of the Abelian
projection without $\vec{X}_{\mu}$
\bea
{\cal L}_{0}=-\dfrac{1}{4}\hat{F}_{\mu\nu}^2
+ \lambda(\hat{n}^2
-1). \label{lag}
\eea
As we have emphasized the theory has
a full SU(2) gauge invariance. More importantly the restricted
gauge theory describes the dual dynamics of QCD with the dynamical
degrees of the maximal Abelian subgroup U(1) as the electric
component and the topological degrees of SU(2) as the magnetic
component~\cite{cho1,cho2}.  In the conventional analysis the
topological degrees have often been neglected because one could
always make $\n$ trivial and remove it from the theory by a gauge
transformation, at least sectionwise locally.  However notice
that, if one includes the topologically non-trivial sectors into
the theory, one can not neglect $\n$ because it can not be
removed by a smooth gauge transformation. In fact it becomes
dynamical when one quantizes the theory \cite{cho00,cho01}.

Obviously the restricted QCD is a constrained system,
because the topological field $\n$ is constrained.
So we have to decide how to deal with the constraint
$\hat{n}^{2}=1$ when we quantize the theory.
In general we could impose the constraint
strictly throughout the quantization, requiring the quantum
fluctuation to respect the constraint. Or we could relax the
constraint, pretending that the quantum fluctuation need not
respect the constraint. In this paper we take the second attitude,
and quantize the restricted QCD treating it as a system with a
second-class constraint. With this understanding we are ready to
derive the first-class Hamiltonian for the restricted QCD in the
framework of the improved Dirac quantization
scheme~\cite{BFT,pr01}, and derive the first-class Hamiltonian for
the restricted QCD.


\section{Constraints and Dirac commutators}
\setcounter{equation}{0}
\renewcommand{\theequation}{\arabic{section}.\arabic{equation}}


In this section, we apply the Dirac quantization
scheme~\cite{dirac} to the restricted QCD with the Lagrangian
(\ref{lag}),
\bea
L_{0}=\int{\rm
d}^{3}x~\left[-\dfrac{1}{4} \hat{F}_{\mu\nu}^2 \right] = \int{\rm
d}^{3}x~\left[-\dfrac{1}{4}G_{\mu\nu}^2 \right], \label{laggg}\nn\\
G_{\mu\nu}=F_{\mu\nu}+H_{\mu\nu}, \label{gmunu}~~~~~~~~~~~~~~~~~~
\eea
regarding it as a second-class constraint system with the
constraint,
\begin{equation}
\Omega_{1}=n^{a}n^{a}-1=\hat{n}^{2}-1\approx 0. \label{c1}
\end{equation}
By performing the Legendre transformation, one can obtain the
momenta conjugate to the fields $A^{\mu}$
\bea
\Omega_{1}^{A}=\Pi^{0}=0,~~~ \Pi^{i}=G^{io}, \label{cojm}
\eea
to yield the canonical Hamiltonian,
\begin{equation}
H=\int{\rm d}^{3}x~\left[\frac{1}{2}\Pi_{i}\Pi_{i}
+\frac{1}{4}G_{ij}^{2}-A_{0}\partial_{i}\Pi_{i}\right]. \label{hc}
\end{equation}
Notice that here one can
explicitly include the constraint term $\lambda\Omega_{1}$ in the
Hamiltonian (\ref{hc}) to yield the primary constraint
$\Omega_{1}^{\lambda} =\pi_{\lambda}=0$ with the momentum
$\pi_{\lambda}$ conjugate to $\lambda$ and, after its time
evolution, the secondary constraint $\Omega_{2}^{\lambda}$. Since
these constraints $\Omega_{i}^{\lambda}$ satisfy a trivial
first-class algebra, we will treat the geometric constraint
$\Omega_{1}$ as usual without explicit inclusion of the
$\lambda\Omega_{1}$ term in the canonical Hamiltonian (\ref{hc}).
With this understanding the time evolution of
the constraint $\Omega_{1}^{A}$ with the
above Hamiltonian yields an additional secondary constraint
\bea
\Omega_{2}^{A}=\partial_{i}\Pi_{i}=0. \label{c2a}
\eea
Note that, the above two constraints $\Omega_{k}^{A}$ are
first-class which satisfy the constraint Lie algebra
\bea
\{\Omega_{k}^{A}(x),\Omega_{k^{\prime}}^{A}(y)\}=0.
\eea

On the other hand, the time evolution of the geometrical
constraint $\Omega_{1}$ with the Hamiltonian (\ref{hc}) yields
\bea
\{\Omega_{1},H\}=-\frac{2g}{c}\hat{n}\cdot\hat{\pi},
\label{comm1h}
~~~~~~~c=\dfrac{1}{g}\hat{n}^{2}(\partial_{i}\hat{n})^2,
\eea
where $\hat{\pi}$ is the momenta conjugate to $\hat{n}$ given by
\bea
\hat{\pi}=\frac{1}{g}(\hat{n}\times\partial_{i}\hat{n})\Pi_{i}.
\label{pia}
\eea
The Poisson bracket (\ref{comm1h}) then yields another secondary
constraint
\bea \Omega_{2}=\hat{n}\cdot\hat{\pi}\approx 0, \label{omega2} \eea
which, together with $\Omega_{1}$, forms a
second-class constraint algebra
\begin{equation}
\Delta_{kk^{\prime}}(x,y)=\{\Omega_{k}(x),\Omega_{k^{\prime}}(y)\}
=2\epsilon_{kk^{\prime}}\hat{n}^{2}\delta(x-y)  \label{delta}
\end{equation}
with $\epsilon_{12}=-\epsilon_{21}=1$. Note that they are involutive with the
above first-class constraints $\Omega_{k}^{A}$
\bea
\{\Omega_{k}^{A}(x),\Omega_{k^{\prime}}(y)\}=0.
\eea
After some algebraic manipulation, we construct the Poisson
bracket
\begin{eqnarray}
\{n^{a}(x),n^{b}(y)\}&=&\{\pi^{a}(x),\pi^{b}(y)\}
=\{A_{\mu}(x),A_{\nu}(y)\}=\{\Pi_{\mu}(x),\Pi_{\nu}(y)\}=0,\nonumber\\
\{n^{a}(x),\pi^{b}(y)\}&=&\delta^{ab}\delta(x-y),  \nonumber \\
\{A_{\mu}(x),\Pi_{\nu}(y)\}&=& g_{\mu \nu}\delta(x-y), \nonumber\\
\{n^{a}(x),A_{\mu}(y)\}&=&\{n^{a}(x),\Pi_{0}(y)\}=\{\pi^{a}(x),A_{0}(y)\}
=\{\pi^{a}(x),\Pi_{0}(y)\}=0,\nonumber\\
\{n^{a}(x),\Pi_{i}(y)\}&=&\frac{1}{c}\epsilon^{abc}n^{b}\partial_{i}n^{c}
\delta(x-y),\nonumber\\
\{\pi^{a}(x),A_{i}(y)\}&=&\frac{1}{g}\epsilon^{abc}n^{b}\partial_{i}n^{c}
\delta(x-y),\nonumber\\
\{\pi^{a}(x),\Pi_{i}(y)\}&=&-\frac{2}{gc}n^{a}(\partial_{i}\hat{n}\cdot
\partial_{j}\hat{n})\Pi_{j}\delta(x-y)+{\rm total~derivative}.
\label{commst}
\end{eqnarray}
Here it is amusing to note that, even though the derivation of the
secondary constraint (\ref{omega2}) and the Poisson algebra
(\ref{commst}) in our theory are quite nontrivial, the set of the
constraints $\Omega_{i}$ are exactly the same as those of the O(3)
nonlinear sigma model~\cite{o3}, where one can easily construct
the constraints.

With the Poisson bracket we can now construct the Dirac brackets
defined as
\bea
\{A(x),B(y)\}_{D} = \{A(x),B(y)\}~~~~~~~~~~~ \nonumber\\
- \int d^2z d^2 z^{\prime} \{A(x),\Omega_{k}(z)\}\Delta^{k
k^{\prime}} \{\Omega_{k^{\prime}}(z^{\prime}),B(y)\},
\eea
where $\Delta^{k k^{\prime}}$ is the inverse of $\Delta_{k
k^{\prime}}$ defined by (\ref{delta}). With this we obtain the Dirac
brackets
\begin{eqnarray}
\{n^{a}(x),\pi^{b}(y)\}_{D}&=&(\delta_{ab}
-\frac{n^{a}n^{b}}{\hat{n}^{2}})\delta(x-y),  \nonumber \\
\{\pi^{a}(x),\pi^{b}(y)\}_{D}&=&\frac{1}{\hat{n}^{2}}
(n^{b}\pi^{a}-n^{a}\pi^{b})\delta (x-y),\nonumber\\
\{\pi^{a}(x),\Pi_{i}(y)\}_{D}&=&-\frac{1}{gc}n^{a}(\partial_{i}\hat{n}\cdot
\partial_{j}\hat{n})\Pi_{j}\delta(x-y)+{\rm total~derivative},\nonumber\\
\{A(x),B(y)\}_{D}&=&\{A(x),B(y)\},~~~~{\rm for~other~commutators.}
\label{dcommst}
\end{eqnarray}
Here note that the structure of the Dirac commutator
$\{\pi^{a}(x),\Pi_{i}(y)\}_{D}$ is nontrivial as much as that of
$\{\pi^{a}(x), \pi^{b}(y)\}_{D}$, since $\Pi_{i}$ is closely
related to $\hat{\pi}$ as shown in Eq. (\ref{pia}).


\section{First-class Hamiltonian}
\setcounter{equation}{0}
\renewcommand{\theequation}{\arabic{section}.\arabic{equation}}


Following the improved Dirac quantization scheme~\cite{BFT,pr01}
which systematically converts the second-class constraints into
the first-class ones, we introduce two St\"uckelberg fields
$(\theta,\pi_{\theta})$ with the Poisson brackets
$$
\{\theta(x), \pi_{\theta}(y)\}=\delta(x-y),
$$
to obtain the first-class constraints as follows
\begin{eqnarray}
\tilde{\Omega}_{1}=\Omega_{1}+2\theta,~  \nonumber \\
\tilde{\Omega}_{2}=\Omega_{2}-\hat{n}^{2}\pi_{\theta},
\label{1stconst}
\end{eqnarray}
which yield a strongly involutive first-class Lie algebra
\bea
\{\tilde{\Omega}_{k}(x),\tilde{\Omega}_{k^{\prime}}(y)\}=0.  \nonumber
\eea
Note that the physical fields $\hat{n}$ are geometrically constrained to fulfill
the modified norm $\hat{n}^{2}=1-2\theta$.

Now, we construct the first-class physical fields
$\tilde{{\cal F}}=(\tilde{n}^{a},\tilde{\pi}^{a})$ corresponding to
the original fields ${\cal F}=(n^{a},\pi^{a})$ demanding that they are
strongly involutive,
\bea
\{\tilde{\Omega}_{k},\tilde{{\cal F}}\}=0. \nonumber
\eea
After some
algebra, we obtain the first-class physical fields as
\begin{eqnarray}
&\tilde{n}^{a}=n^{a}\left(\dfrac{\hat{n}^{2}+2\theta}{\hat{n}^{2}}
\right)^{1/2},~~~
\tilde{\pi}^{a}=\left(\pi^{a}-n^{a}\pi_{\theta}\right)
\left(\dfrac{\hat{n}^{2}}{\hat{n}^{2}+2\theta}\right)^{1/2},
\nonumber\\
&\tilde{A}_{\mu}=A_{\mu},~~~~\tilde{\Pi}_{\mu}=\Pi_{\mu},
\nonumber\\
&\tilde{G}_{ij}=G_{ij}+H_{ij}
\left[\left(\dfrac{\hat{n}^{2}+2\theta}
{\hat{n}^{2}}\right)^{3/2}-1\right],~~~ \tilde{G}_{0i}=G_{0i}.
\label{pitilde}
\end{eqnarray}
Since any functional of the first-class fields $\tilde{{\cal F}}$ is also
first-class~\cite{kpr}, we easily construct a first-class Hamiltonian in
terms of the above first-class physical variables omitting infinite iteration procedure
to arrive at
\begin{equation}
\tilde{H}=\int {\rm
d}^{3}x~\left[\frac{1}{2}\tilde{\Pi}_{i}\tilde{\Pi}_{i}
+\frac{1}{4}\tilde{G}_{ij}^{2}-\tilde{A}_{0}\partial_{i}\tilde{\Pi}_{i}\right].
\label{htilde}
\end{equation}
We then directly rewrite this Hamiltonian in terms of the original
as well as St\"uckelberg fields to obtain
\begin{eqnarray}
\tilde{H}=\int {\rm d}^{3}x \left[\frac{1}{2}\Pi_{i}\Pi_{i}+
\frac{1}{4}(G_{ij}+H_{ij}D)^{2}-A_{0}\partial_{i}\Pi_{i}\right],
\label{hct}
\end{eqnarray}
where $D=R^{3/2}-1$ with the rescaling factor $R$
\bea
R=\frac{\hat{n}^{2}+2\theta}{\hat{n}^{2}}. \nn
\eea
Note that $\tilde{H}$
is strongly involutive with the first-class constraints,
\bea
\{\tilde{\Omega}_{k},\tilde{H}\}=0. \nn
\eea

Next, we consider the partition function of the theory in order to present
the Lagrangian corresponding to the first-class Hamiltonian $\tilde{H}$ in Eq.(\ref{hct}).
The starting partition function in the phase space is then given by
the Faddeev-Senjanovic formula \cite{faddev} as follows
\begin{eqnarray}
Z=N\int {\cal D}n^{a}{\cal D}\pi^{a}{\cal D}A_{\mu}{\cal
D}\Pi_{\mu} {\cal D}\theta{\cal D}\pi_{\theta}
\prod_{i,j=1}^{2}\delta(\tilde{\Omega}_i)\delta(\Gamma_j)
\nonumber\\
\cdot\det|\{\tilde {\Omega}_i,\Gamma_j\}|\exp \left[{i\int {\rm d}t L}\right],~~~~~~~~~~~~~~  \nonumber\\
L=\int {\rm
d}^{3}x~\left[\pi^{a}\dot{n}^{a}+\Pi_{\mu}\dot{A}_{\mu}+\pi_{\theta}\dot{\theta}
-\tilde{H}\right],~~~~~~
\end{eqnarray}
where the gauge fixing conditions $\Gamma_{i}$ are chosen so that the
determinant occurring in the functional measure is nonvanishing.

Now, exponentiating the delta function $\delta
(\tilde{\Omega}_{2})$ with
\bea \delta (\tilde{\Omega}_{2})=\int
{\cal D}\xi \exp\left[{i\int {\rm d}t~\xi
\tilde{\Omega}_{2}}\right],
\eea
and performing the integration over $\pi^{a}$,
$\Pi_{\mu}$, $\pi_{\theta}$, and $\xi$, we obtain
the following partition function
and Lagrangian
\begin{eqnarray}
Z=N\int {\cal D}n^{a}{\cal D}A_{\mu}{\cal D}\theta
\delta(\tilde{\Omega}_{1})
\prod_{i=1}^{2}\delta(\Gamma_i)\det|\{\tilde{\Omega}_i,\Gamma_j\}|
\exp \left[{i\int
{\rm d}t L}\right],  \label{fca} \nn\\
L=L_{0}+L_{WZ},~~~~~~~~~~~~~~~~~~~~~~~~~~~~~~~~~~~~ \nonumber\\
L_{WZ}=\int{\rm
d}^{3}x~\left[-\frac{1}{4}H_{ij}(2G_{ij}+H_{ij}D)D\right],
\label{zhct}~~~~~~~~~~~~~~~~
\end{eqnarray}
which is the desired first-class Lagrangian corresponding to the first-class
Hamiltonian (\ref{hct}) and is invariant under the following gauge
transformations
\begin{equation}
\delta n^{a}=n^{a}\epsilon,~~~\delta \theta=-(1-2\theta)\epsilon,
~~~\delta A_{\mu}=0.
\label{gaugetrfm}
\end{equation}
Here one notes that, by using the first-class observables
$\tilde{G}_{\mu\nu}$
in Eq. (\ref{pitilde}), the Lagrangian (\ref{zhct}) can be
reshuffled to yield the Lorentz invariant form
\beq
L=\int{\rm d}^{3}x~\left[-\frac{1}{4} \tilde{G}_{\mu\nu} ^2
\right]. \label{zhct2}
\eeq


\section{BRST symmetry}
\setcounter{equation}{0}
\renewcommand{\theequation}{\arabic{section}.\arabic{equation}}


In this section, in order to obtain the effective Lagrangian, we
introduce two canonical sets of ghosts and anti-ghosts together
with auxiliary fields in the framework of the
Batalin-Fradkin-Vilkovisky formalism \cite{bfv,fik}, which is
applicable only to theories with the first-class constraints,
\[
({\cal C}_{i},{\cal P}^{\dag}_{i}),~~({\cal P}_{i}, {\cal
C}^{\dag}_{i}), ~~(N_{i},B_{i}),~~~~(i=1,2)
\]
which satisfy the super-Poisson algebra
\[
\{{\cal C}_{i}(x),{\cal P}^{\dag}_{j}(y)\}=\{{\cal P}_{i}(x),
{\cal C}^{\dag}
_{j}(y)\}=\{N_{i}(x),B_{j}(y)\}=\delta_{ij}\delta(x-y),
\]
where the super-Poisson bracket is defined as
\[
\{A,B\}=\frac{\delta A}{\delta q}|_{r}\frac{\delta B}{\delta p}|_{l}
-(-1)^{\eta_{A}\eta_{B}}\frac{\delta B}{\delta q}|_{r}\frac{\delta A} {%
\delta p}|_{l}.
\]
Here $\eta_{A}$ and $\eta_{B}$ denotes the number of fermions
called ghost number in $A$ and $B$, and the subscript $r$ and $l$
imply right and left derivatives, respectively.
In this formalism, the nilpotent BRST charge $Q$ and the
fermionic gauge fixing function $\Psi$ are given as
\begin{eqnarray}
Q=\int {\rm d}^{3}x~({\cal C}_{i}\tilde{\Omega}_{i}+{\cal
P}_{i}B_{i}),
\nonumber \\
\Psi=\int {\rm d}^{3}x~({\cal C}^{\dag}_{i}\chi_{i}+{\cal
P}^{\dag}_{i}N_{i}).\label{qnpsi}
\end{eqnarray}
Now we choose a gauge
\begin{equation}
\chi_{1}=\Omega_{1},~~~\chi_{2}=\Omega_{2}. \label{gaugefix}
\end{equation}
With this the BRST charge $Q$ and the fermionic gauge fixing
function $\Psi$ satisfy the following relations
\begin{equation}
\{Q,\tilde{H}\}=0,~~Q_{2}=\{Q,Q\}=0,~~\{\{\Psi,Q\},Q\}=0.
\end{equation}
The quantum Lagrangian is then described by
\bea
L=\int {\rm d}^{3}x~(\pi^{a}\dot{n}^{a}+\Pi_{\mu}\dot{A}_{\mu}+\pi_{\theta}\dot{\theta} +B_{2}%
\dot{N}_{2}+{\cal P}^{\dag}_{i}\dot{{\cal C}}_{i}+{\cal
C}^{\dag}_{2} \dot{{\cal P}}_{2})-H_{tot}, \nonumber \\
H_{tot}=\tilde{H}-\{Q,\Psi\},~~~~~~~~~~~~~~~~~~~~~~~~~~~~
\eea
where we have suppressed the term
\bea
\int {\rm d}^{3}x~(B_{1} \dot{N}_{1} +{\cal
C}^{\dag}_{1}\dot{{\cal P}}_{1})=\{Q,\int{\rm d}^{3} x~{\cal
C}^{\dag}_{1} \dot{N}_{1}\}
\eea
by replacing $\chi_{1}$ with $\chi_{1} +\dot{N}_{1}$.

Integrating out $N_{1}$, $B_{1}$, ${\cal C}^{\dag}_{1}$, ${\cal
C}_{1}$, and the momentum fields by performing path integration in the
BRST scheme, we can obtain the Lagrangian of the form
\bea
L=\int{\rm d}^{3}x~\left[-\frac{1}{4}G_{\mu\nu} ^2
-\frac{1}{4}H_{ij}(2G_{ij}+H_{ij}D)D +B\dot{N} +\dot{\cal
C}^{\dag}\dot{{\cal C}}\right],
\eea
where
\bea
N=N_{2},~~~ B=B_{2},~~~ {\cal C}^{\dag}={\cal
C}^{\dag}_{2},~~~ {\cal C}={\cal C}_{2}. \nn
\eea
Identifying
\bea
N=-B+\frac{\dot{\theta}}{1-2\theta}, \nn
\eea
we obtain
\begin{eqnarray}
L=L_{0}+L_{WZ}+L_{gh},~~~~~~~~~~~ \nonumber\\
L_{gh}=\int{\rm d}^{3}x~\left[
-\frac{\partial_{\mu}B\partial^{\mu}{\theta}}{1-2\theta}
+\partial_{\mu}{\cal C}^{\dag}\partial^{\mu}{\cal C}\right],
\label{efflag}
\end{eqnarray}
which is invariant under the BRST transformation
\bea
\delta_{B}n^{a}=\lambda n^{a}{\cal C},~~~
\delta_{B}\theta=-\lambda (1-2\theta){\cal C},~~~
\delta_{B}A_{\mu}=0, \nn \\
\delta_{B}{\cal C}^{\dag}=-\lambda B,~~~ \delta_{B}{\cal C}=0,~~~
\delta_{B}B=0.~~~~~~~~ \label{brsttrfm}
\eea
Now, using the first-class observables $\tilde{G}_{\mu\nu}$ in Eq.
(\ref{pitilde}), we can obtain the Lorentz invariant form of the
Lagrangian (\ref{efflag}) as follows
\bea L=\int{\rm
d}^{3}x~\left[-\frac{1}{4}\tilde{G}_{\mu\nu} ^2
-\frac{\partial_{\mu}B\partial_{\mu}{\theta}}{1-2\theta}
+\partial_{\mu}{\cal C}^{\dag}\partial_{\mu}{\cal
C}\right].\label{lorentzinv}
\eea
Moreover, the above Lagrangian can be rewritten in terms of the
first-class physical fields
\bea Z=N\int {\cal D}\tilde{n}^{a}{\cal D} \tilde{A}_{\mu}{\cal
D}\tilde{\theta} {\cal D} B {\cal D} {\cal C} {\cal D} {\cal
C}^\dag \delta(\tilde{\Omega}_{1}) \exp \left[{i\int {\rm d}t L}
\right],
\label{fcab} \nn\\
L=\int{\rm d}^{3}x~\left[ -\frac{1}{4}\tilde{G}_{\mu\nu} ^2
+\frac{1}{3}\partial_{\mu}B\partial_{\mu}
(\ln\tilde{\theta})+\partial_{\mu}{\cal
C}^{\dag}\partial_{\mu}{\cal C}\right], \label{lagtilb}~~~
\eea
where
\bea
\tilde{\theta} =(1-2\theta)^{3/2}. \nn
\eea
Note that in this form of
Lagrangian (\ref{lagtilb}) the BRST
transformation (\ref{brsttrfm}) is expressed by
\bea
\delta_{B}\tilde{n}^{a}=0,~~~
\delta_{B}\tilde{\theta}=3\lambda\tilde{\theta}{\cal C},~~~
\delta_{B}\tilde{A}_{\mu}=0,\nn\\
\delta_{B}{\cal C}^{\dag}=-\lambda B,~~~ \delta_{B}{\cal C}=0,~~~
\delta_{B}B=0.~ \label{brsttrfm2}
\eea

With the above discussion we notice that the gauge condition
(\ref{gaugefix}) requires $\theta =0$, and effectively removes it
from the Lagrangian (\ref{lorentzinv}). In this case $ {\tilde
A}_{\mu} $, $ \tilde n $, and ${\tilde G}_{\mu \nu}$ becomes
identical to ${\hat A}_\mu$, $ \hat n $, and $G_{\mu \nu}$.
Furthermore the ghost fields $\cal C$ and ${\cal C}^{\dag}$
completely decouple from the gauge field ${\hat A}_\mu$.
So we finally have
\bea Z=N\int {\cal D} n^a {\cal D} A_\mu
\delta({\hat n}^2 - 1) \exp \left[{i\int {\rm d}t L} \right],
\nn \\
L=\int{\rm d}^{3}x~ \left[-\frac{1}{4} {\hat F}_{\mu\nu}^2 \right]
, \label{znl} ~~~~~~~~~~~~~~~
\eea
This shows that the improved Dirac quantization rule and the direct
quantization with the constraint ${\hat n}^2 =1$ produce an
identical result, at least in the gauge (\ref{gaugefix}).
An interesting problem is to prove the equivalence of the
two approaches in an arbitrary gauge.

It should be remarked that, in order to complete
the quantization of the restricted
QCD described by the above Lagrangian (\ref{znl}), we still have to fix
the gauge degrees of SU(2) even after
the gauge fixing (\ref{gaugefix}). For example
we may choose the Lorentz gauge as the SU(2)
gauge fixing condition \cite{cho01}
\begin{equation}
\partial_{\mu} {\hat{A}}_{\mu}=0,
\label{gaugeconatil}
\end{equation}
which can be splitted into two parts
\bea
\partial_{\mu} A_{\mu}=0,
\label{gau1}~~~~~~~~ \nn \\
\hat n \times\partial^2 \hat n - g A_{\mu}\partial_{\mu}
\hat n =0. \label{gau2}
\eea
In this case the corresponding Faddeev-Popov determinant is given
by
\bea M_{ab} = \dfrac{\delta (\partial_\mu
{\hat{A}}_\mu)_a}{\delta \alpha^b} = (\partial_\mu
{\hat{D}}_\mu)_{ab}, \label{determinant}
\eea
from which we obtain
\bea
Z=N\int {\cal D} n^a {\cal D} A_\mu
{\cal D} {\vec c}~^{\dag} {\cal D} {\vec c}
\delta({\hat n}^2 - 1) \exp \left[{i\int {\rm d}t L_{eff}} \right], \nn \\
L_{eff}=\int{\rm d}^{3}x~\left[-\frac{1}{4} {\hat F}_{\mu\nu}^2 \right.
~~~~~~~~~~~~~~~~~~~~~~~~~~~\nn \\
- \left.\dfrac{1}{2\xi} \left( (\partial_\mu A_\mu)^2
+(\hat{n}\times\partial^2 \hat{n}
- gA_{\mu}\partial_{\mu} \hat{n})^{2} \right)
+{\vec c}~^{\dag} \cdot \partial_{\mu}{\hat D}_{\mu} \vec{c} \right],
\eea
where ${\vec c}$ and ${\vec c}~^{\dag}$ are the ghost fields
associated with the determinant (\ref{determinant})
\cite{cho01}.


\section{Conclusions}
\setcounter{equation}{0}
\renewcommand{\theequation}{\arabic{section}.\arabic{equation}}


In this paper, we have discussed the restricted QCD, which
possesses the second-class constraints, to construct
the first class Hamiltonian and first-class effective Lagrangian
in the framework of the improved Dirac quantization scheme.
Furthermore, with the first-class effective Lagrangian, we have
constructed the BRST symmetry of the restricted QCD.

The lesson that we learn from the discussion is that the improved
Dirac quantization method guarantees that, in the restricted QCD,
the second-class constraint (\ref{c1}) can be treated as if it
is a first-class constraint. This tells that we no longer have to worry about
whether the constraint of the topological field $\n$ is of the
first-class or second-class, when we quantize the restricted QCD.
This is remarkable. This has been demonstrated by
a proper choice of the gauge (\ref{gaugefix}). The remaining problem is
to establish the equivalence of the two approaches in an arbitrary gauge
where $\theta$ is not zero, which is worth further
investigation.

\vskip 1.0cm Two of the authors (YMC and STH) would like to thank
the hospitality of J. Michelsson and E. Langman of Department
of Theoretical Physics at Royal Institute of Technology and
A. Niemi of Department of Theoretical Physics at Uppsala
University, where this work has been initiated. The work of YMC
and JHK is supported in part by the ABRL Project of Korea Research 
Foundation (KRF-R14-2003-012-01002-0), and the work of STH and 
YJP was supported by the Korea Research Foundation Grant 
KRF-2006-331-C00071 and KRF-2005-015-C00105 funded by Korean
Government.


\begin{thebibliography}{99}
\bibitem{nambu} Y. Nambu, Phys. Rev. D {\bf 10}, 4262 (1974);
S. Mandelstam, Phys. Rep. {\bf 23}, 245 (1976); A. Polyakov, Nucl.
Phys. B {\bf 120}, 429 (1977).
\bibitem{cho1} Y.M. Cho, Phys. Rev. D {\bf 21}, 1080 (1980); J. Korean Phys.
Soc. {\bf 17}, 266 (1984).
\bibitem{hooft} G. 't Hooft, Nucl. Phys. B {\bf 190}, 455 (1981).
\bibitem{cho2} Y.M. Cho, Phys. Rev. Lett. {\bf 46}, 302 (1981);
Phys. Rev. D {\bf 23}, 2415 (1981).
\bibitem{cho00} Y.M. Cho, Phys. Rev. D {\bf 62}, 074009 (2000).
\bibitem{dirac}  P.A.M. Dirac, {\it Lectures in Quantum Mechanics} (Yeshiva
University, New York, 1964).
\bibitem{BFT} I.A. Batalin and E.S. Fradkin, Phys. Lett. B {\bf 180}, 157 (1986);
Nucl. Phys. B {\bf 279}, 514 (1987); I.A. Batalin and I.V. Tyutin,
Int. J. Mod. Phys. A {\bf 6}, 3255 (1991).
\bibitem{pr01} S.T. Hong and Y.J. Park, Phys. Rep. {\bf 358}, 143 (2002).
\bibitem{sav} G. K. Savvidy, Phys. Lett. B {\bf 71}, 133 (1977);
N. K. Nielsen and P. Olesen, Nucl. Phys. B {\bf 144}, 485 (1978).
\bibitem{yil} A. Yildiz and P. Cox, Phys. Rev. D {\bf 21}, 1095 (1980);
M. Claudson, A. Yilditz, and P. Cox, Phys. Rev. D {\bf 22}, 2022
(1980).
\bibitem{prd02} Y.M. Cho, H.W. Lee, and D.G. Pak, Phys. Lett. B {\bf 525}, 347 (2002);
Y.M. Cho and D.G. Pak, Phys. Rev. D {\bf 65}, 074027 (2002).
\bibitem{jhep} Y.M. Cho, D.G. Pak, and M. Walker, JHEP {\bf 05}, 073 (2004);
Y.M. Cho and M.L. Walker, Mod. Phys. Lett. A {\bf 19}, 2707
(2004); Y.M. Cho, J.H. Kim, and D.G. Pak, Mod. Phys. Lett. A {\bf
21}, in press.
\bibitem{faddeev1} L. Faddeev and A. Niemi, Nature {\bf 387}, 58 (1997);
R. Battye and P. Sutcliffe, Phys. Rev. Lett. {\bf 81}, 4798
(1998).
\bibitem{faddeev2} L. Faddeev and A. Niemi, Phys. Rev. Lett. {\bf 82}, 1624 (1999);
Phys. Lett. B {\bf 449}, 214 (1999).
\bibitem{niemi99} E. Langman and A. Niemi, Phys. Lett. B {\bf 463}, 252 (1999);
S. Shabanov, Phys. Lett. B {\bf 458}, 322 (1999); B {\bf 463}, 263
(1999); H. Gies, Phys. Rev. D {\bf 63}, 125023 (2001).
\bibitem{cho01} W.S. Bae, Y.M. Cho, and S.W. Kimm, Phys. Rev. D {\bf 65}, 025005 (2001).
\bibitem{prl01} Y.M. Cho, Phys. Rev. Lett. {\bf 87}, 252001 (2001);
Y.M. Cho, Phys. Lett. B {\bf 603}, 88 (2004).
\bibitem{plb05} Y.M. Cho, Phys. Lett. B {\bf 616}, 101 (2005).
\bibitem{plb06} P. van Baal and A. Wipf, Phys. Lett. B {\bf 515}, 181 (2001);
Y.M. Cho, hep-th/0409246, Phys. Lett. B, in press.
\bibitem{wu} T.T. Wu and C.N. Yang, Phys. Rev. D {\bf 12}, 3845 (1975).
\bibitem{choprl80} Y.M. Cho, Phys. Rev. Lett. {\bf 44}, 1115 (1980);
Phys. Lett. B {\bf 115}, 125 (1982); Y.D. Kim, I.G. Koh, and Y.J.
Park, Phys. Rev. D {\bf 25}, 587 (1982); W.S. l'Yi, Y.J. Park,
I.G. Koh, and Y. Kim, Phys. Rev. Lett. {\bf 49}, 1229 (1982).
\bibitem{bpst}A. Belavin, A. Polyakov, A. Schwartz, and Y. Tyupkin, Phys.
Lett. B {\bf 59}, 85 (1975); G. 't Hooft, Phys. Rev. Lett. {\bf
37}, 8 (1976).
\bibitem{cho79} Y.M. Cho, Phys. Lett. B {\bf 81}, 25 (1979).
\bibitem{sk2}  W. Oliveira and J.A. Neto, Int. J. Mod. Phys. A {\bf
12}, 4895 (1997); S.T. Hong, Y.W. Kim, and Y.J. Park, Phys. Rev. D
{\bf 59}, 114026 (1999); S.T. Hong and Y.J. Park, Phys. Rev. D
{\bf 63}, 054018 (2001).
\bibitem{o3} S.T. Hong, W.T. Kim, and Y.J. Park, Phys. Rev. D {\bf 60}, 125005 (1999).
\bibitem{hong05prd} S.T. Hong and A.J. Niemi, Phys. Rev. D {\bf 72}, 127701 (2005).
\bibitem{hong00q}  D.K. Hong, S.T. Hong, and Y.J. Park, Phys. Lett. B {\bf 499}, 125 (2001).
\bibitem{hong00db}  S.T. Hong, W.T. Kim, Y.J. Park, and M.S. Yoon, Phys. Rev. D {\bf 62}, 085010 (2000).
\bibitem{zucc} R. Zucchini, Int. J. Geom. Meth. Mod. Phys. {\bf 1}, 813 (2004).
\bibitem{kondo} K. Kondo, Phys. Lett. B {\bf 600}, 287 (2004); Int. J. Mod. Phys. A {\bf 20}, 4609 (2005);
K. Kondo, T. Murakami, and T. Shinohara, Euro Phys. J. C {\bf 42},
475 (2005); S. Kato et al, Phys. Lett. B {\bf 71}, 133 (2006).
\bibitem{kpr} W.T. Kim, Y.W. Kim, M.I. Park, Y.J. Park, and S.J. Yoon, J. Phys. G {\bf 23}, 325 (1997);
Y.W. Kim and K. D. Rothe, Nucl. Phys. B {\bf 510}, 511 (1998).
\bibitem{faddev} L.D. Faddeev, Theor. Math. Phys. {\bf 1}, 1 (1970);
P. Senjanovich, Ann. Phys. {\bf 100}, 277 (1976).
\bibitem{bfv} E.S. Fradkin, G.A. Vilkovisky, Phys. Lett. B {\bf 55}, 224 (1975);
M. Henneaux, Phys. Rep. {\bf 126}, 1 (1985).
\bibitem{fik} T. Fujiwara, Y. Igarashi and J. Kubo, Nucl. Phys. B {\bf 341}, 695 (1990);
Y.W. Kim, S.K. Kim, W. T. Kim, Y.J. Park, K. Y. Kim, and Y. Kim,
Phys. Rev. D {\bf 46}, 4574 (1992); R. Banerjee, H.J. Rothe, and
K.D. Rothe, Phys. Rev. D {\bf 49}, 5438 (1994); C. Bizdadea and
S.O. Saliu, Nucl. Phys. B {\bf 456}, 473 (1995).
\end{thebibliography}
\end{document}